\documentclass[11pt]{article}
 \pdfoutput=1
 
 \usepackage{jheppub}
 
 \usepackage[T1]{fontenc}

\usepackage{graphicx}
\usepackage{amssymb}
\usepackage{amsmath}

\usepackage{hyperref}

\usepackage{cancel}

\newcommand*{\TextVCenter}[1]{%
  \text{$\vcenter{\hbox{#1}}$}%
}

\usepackage{accents}
\newlength{\dhatheight}
\newcommand{\doublehat}[1]{%
    \settoheight{\dhatheight}{\ensuremath{\hat{#1}}}%
    \addtolength{\dhatheight}{-0.3ex}%
    \hat{\vphantom{\rule{1pt}{\dhatheight}}%
    \smash{\hat{#1}}}}

\newcommand{\abs}[1]{\left\lvert #1 \right\rvert}
\newcommand{\tr}{\operatorname{tr}}

\title{\huge Constraining the Higgs trilinear coupling from an $SU(2)$ quadruplet with bounded-from-below conditions}
\author{K. Kannike}

\affiliation[]{National Institute of Chemical Physics and Biophysics, R\"avala 10, 10143, Tallinn, Estonia}

\emailAdd{kristjan.kannike@cern.ch}

\abstract{Integrating out a heavy scalar can cause the Higgs trilinear coupling to deviate from its Standard Model value: a good example is provided by an $SU(2)$ quadruplet. Constraints on the full theory, however, can limit the size of the deviation. We show that the bounded-from-below conditions for the Standard Model extended by an $SU(2)$ quadruplet strongly constrain the $\mathbb{Z}_{2}$-breaking Higgs portal and can bound the Higgs trilinear coupling close to its Standard Model value. For TeV-scale quadruplet masses in models with custodial symmetry violation, these constraints can be a few times stronger than constraints from electroweak precision measurements. For the custodial quadruplet, these are the strongest theoretical constraints available.}

\begin{document}
\maketitle

\flushbottom

\section{Introduction}
\label{sec:introduction}

The discovery of the Higgs boson a decade ago was a spectacular confirmation of the validity of the Standard Model (SM) \cite{CMS:2012qbp,ATLAS:2012yve}. No other new elementary particles have been found at the LHC. Heavy undiscovered particles, however, can indirectly influence the couplings of the Higgs boson, especially its self-coupling and the associated Higgs trilinear coupling measured at colliders. In particular, the presence of new scalars can change the trilinear coupling significantly (see e.g. \cite{Bahl:2022jnx}). At present, the Higgs trilinear coupling is very loosely bounded: its ratio to its SM value, $\kappa_{\lambda}$, is constrained by $-1.24 < \kappa_{\lambda} < 6.49$ at the CMS \cite{Collaboration:2022yf} and $-0.4 < \kappa_{\lambda} < 6.3$ at ATLAS \cite{ATLAS:2022jtk} at $95\%$ confidence level. The High Luminosity LHC is expected to bring more precision with an estimated $0.1 < \kappa_{\lambda} < 2.3$ \cite{Cepeda:2019klc}, while the ILC or the FCC-\emph{hh} can constrain it to within $10\%$ of its SM value \cite{Cepeda:2019klc,Fujii:2017vwa,Goncalves:2018qas}. Before these higher precision measurements, our knowledge of the full scalar field content and the scalar potential remains scanty.

The minimal addition to the SM Higgs doublet is given by a single scalar multiplet $\phi$. A representative example of such a new scalar is provided by an $SU(2)$ quadruplet \cite{AbdusSalam:2013eya,Dawson:2017vgm}. Depending on its hypercharge, it can have different interactions with the Higgs doublet: the phenomenologically interesting possibilities are $Y = 1/2$ or $Y = 3/2$. The heavy quadruplet can be integrated out, leaving behind effective contributions to the SM \cite{Durieux:2022hbu}. In particular, the $\mathbb{Z}_{2}$-breaking Higgs-quadruplet coupling $\hat{\lambda}_{H\phi}$ will induce the sixth order $\abs{H}^{6}$ effective coupling to which the deviation in the Higgs trilinear coupling is proportional. The models with either $Y = 1/2$ or $Y = 3/2$ quadruplet alone violate custodial symmetry and are restricted by electroweak precision measurements, while a combination, the custodial quadruplet, has no such restrictions \cite{Durieux:2022hbu}.

The full theory must obey additional constraints, such as perturbative unitarity and bounded-from-below conditions in the presence of the quadruplet. In the full theory, a sizeable value of the $\mathbb{Z}_2$-breaking $\hat{\lambda}_{H\phi}$ coupling can make the potential unbounded from below, unless balanced by other large, $\mathbb{Z}_2$-preserving, scalar couplings.

Our goal is to calculate the bounded-from-below conditions for the scalar potentials with $Y = 1/2$ or $Y = 3/2$ quadruplet and use them to constrain the $\hat{\lambda}_{H\phi}$ coupling. Additional constraints are provided by perturbative unitarity. The bounded-from-below conditions are derived using copositivity constraints \cite{Kannike:2012pe} and positivity of quartic polynomials \cite{Kannike:2016fmd} on the gauge orbit space of gauge invariants \cite{Abud:1981tf,Kim:1981xu,Abud:1983id,Kim:1983mc}. We use the $P$-matrix \cite{Talamini:2006wd,Abud:1983id,Abud:1981tf} and birdtrack \cite{Cvitanovic:1976am,Cvitanovic:2008zz} methods to find the orbit space. We show that the bounded-from-below conditions can constrain the Higgs couplings several times more strongly than electroweak precision measurements for quadruplet masses above a couple of TeV (see \cite{Durieux:2022hbu}) in models that break custodial symmetry. For the custodial quadruplet, the bounded-from-below conditions can be the strongest constraint, depending on the value of the Higgs effective quartic coupling.

We write the scalar potential of $Y = 1/2$ and $Y = 3/2$ quadruplets in section~\ref{sec:potential}. The orbit space is determined in section~\ref{sec:orbit:space}. Bounded-from-below conditions for the scalar potential are derived in section~\ref{sec:bfb} and perturbative unitarity constraints in section~\ref{sec:pert:unit}. The results are given in section~\ref{sec:results} and conclusions in section~\ref{sec:conclusions}. A birdtrack derivation of the $P$-matrix is given in Appendix~\ref{sec:pmatrix}.

\section{Scalar potential}
\label{sec:potential}

We consider the most general scalar potential of the Higgs doublet $H$ and an $SU(2)$ quadruplet $\phi$ with hypercharge $Y = 1/2$ or $Y = 3/2$. We take the Higgs doublet
\begin{equation}
  H = 
  \begin{pmatrix}
    0
    \\
    \frac{h}{\sqrt{2}}
  \end{pmatrix}
\end{equation}
in the unitary gauge. The $SU(2)_{L}$ quadruplet is given by a symmetric tensor $\phi_{ijk} = \phi_{(ijk)}$. 
In the case of $Y = 3/2$, for example, its four components are given in terms of the four canonically normalised propagating degrees of freedom as $\phi_{111} = \phi^{3+}$, $\phi_{112} = \phi_{211} = \phi_{121} = \phi^{++}/\sqrt{3}$, $\phi_{122} = \phi_{221} = \phi_{212} = \phi^{+}/\sqrt{3}$ and $\phi_{222} = \phi^{0}$.

The scalar potential is the sum of the part invariant under the $\mathbb{Z}_{2}$ symmetry $\phi \to -\phi$,
\begin{equation}
\begin{split}
  V_{\mathbb{Z}_{2}} &= \mu_{H}^{2} H^{*i} H_{i} + \mu_{\phi}^{2} \phi^{*ijk} \phi_{ijk} + \lambda_{H} (H^{*i} H_{i})^{2} 
  + \lambda_{\phi} (\phi^{*ijk} \phi_{ijk})^{2} 
  \\
  &+ \lambda'_{\phi} \phi^{*ijk} \phi_{ijn} \phi^{*lmn} \phi_{lmk} 
  + \lambda_{H\phi} (H^{*l} H_{l}) (\phi^{*ijk} \phi_{ijk})
  + \lambda'_{H\phi} H^{*i} H_{k} \phi^{*ljk} \phi_{lji},
\end{split}
\end{equation}
and a $\mathbb{Z}_{2}$-breaking part, which for $Y = 1/2$ is given by%
\footnote{Thanks to Luis Lavoura for pointing out the $\doublehat{\lambda}_{H\phi}$ interaction term that was missing in some of the literature. However, this term does not contribute to the Higgs trilinear coupling. It is also easy to show that it can be made negative simultaneously with the $\hat{\lambda}_{H\phi}$ term. In order to to find the maximum allowed parameter space for the Higgs trilinear coupling, we set $\doublehat{\lambda}_{H\phi} = 0$ in the rest of the paper. Finding the full orbit space and bounded-from-below conditions involving this term is beyond the scope of the current paper.}
\begin{equation}
\begin{split}
  V_{\cancel{\mathbb{Z}_{2}}} &= \hat{\lambda}_{H\phi} (\phi^{*ijk} H_{i} H_{j} \epsilon_{kl} H^{*l} + \phi_{ijk} H^{*i} H^{*j} \epsilon^{lk} H_{l})
  \\
  &+ \doublehat{\lambda}_{H\phi}  (\phi^{ijk} \epsilon_{jm} \epsilon_{kn} \phi^{* lmn} \phi_{ilp} \epsilon^{pq} H_{q} 
  + \phi_{ijk} \epsilon^{mj} \epsilon^{nk} \phi_{lmn} \phi^{* ilp} \epsilon_{qp} H^{*q}),
\end{split}
\label{eq:V:Z:2:breaking:Y:1:2}
\end{equation}
and for $Y = 3/2$ by
\begin{equation}
  V_{\cancel{\mathbb{Z}_{2}}} = \hat{\lambda}_{H\phi} (\phi^{*ijk} H_{i} H_{j} H_{k} + \phi_{ijk} H^{*i} H^{*j} H^{*k}).
\label{eq:V:Z:2:breaking:Y:3:2}
\end{equation}
We have taken $\hat{\lambda}_{H\phi}$ and $\doublehat{\lambda}_{H\phi}$ to be real without loss of generality. Note that other works may use different notation for the $\hat{\lambda}_{H\phi}$ coupling, such as $\lambda$ or $\lambda_{\phi}$, possibly with different normalisation. With respect to the $\lambda_{\tilde{\Phi}}$ of the $Y = 3/2$ quadruplet of Ref.~\cite{Durieux:2022hbu}, in particular, we have $\hat{\lambda}_{H\phi} = \lambda_{\tilde{\Phi}}/\sqrt{3}$, while for the $Y = 1/2$ quadruplet the normalisation is the same as ours.

When the quadruplet is integrated out, it will generate a deviation $\delta_{h^{3}}$ in the trilinear Higgs coupling, given by $\delta_{h^{3}} = \kappa_{\lambda} - 1$. At dimension $d = 6$, one has \cite{Durieux:2022hbu}
\begin{equation}
  \delta_{h^{3}} = \frac{c_{6}}{M_{\phi}^{2} G_{F}^{2} m_{h}^{2}},
\end{equation}
where $G_{F}$ is the Fermi constant, $m_{h}$ is the SM Higgs boson mass, and 
\begin{equation}
  c_{6} = -\frac{\hat{\lambda}_{H\phi}^{2}}{3} \text{ for $Y = \frac{1}{2}$ \quad and \quad }
  c_{6} = -\frac{(\sqrt{3} \hat{\lambda}_{H\phi})^{2}}{3} \text{ for $Y = \frac{3}{2}$}.
  \label{eq:c:6}
\end{equation}
The custodial quadruplet, transforming as a $(\mathbf{4}, \mathbf{4})$ under the custodial $SO(4) \simeq SU(2)_{L} \times SU(2)_{R}$ symmetry, decomposes into a $Y = 1/2$ and a $Y = 3/2$ quadruplet with $\hat{\lambda}_{H\phi}^{Y=1/2} = \sqrt{3} \hat{\lambda}_{H\phi}^{Y=3/2}$ and equal masses. In the custodial case, the contributions in Eq.~\eqref{eq:c:6} are therefore equal  and sum up.

\section{Orbit space}
\label{sec:orbit:space}

\subsection{Orbit space and $P$-matrix formalism}
\label{sec:orbit:space:P:form}

Under a gauge transformation $T(\theta)$ in the gauge group $G$, the components of a constant scalar field configuration $\Phi$ (e.g. a vacuum expectation value) will rotate among themselves through a gauge orbit as $\Phi \to \Phi_{\theta} = T(\theta) \Phi$. Since the scalar potential $V(\Phi)$ is gauge-invariant, its value does not change under the transformation. In particular, for a unitary group all the states $\Phi_{\theta}$ have the same norm $\Phi^{*} \Phi$.

For a compact group, all gauge-invariant polynomials of scalar fields can be given as combinations of elements of a finite polynomial basis (minimal integrity basis) $p_a$ with $a = 1, \ldots, q$. The polynomial basis always includes the norms of fields. Therefore, we can write the scalar potential -- which is gauge-invariant -- as a function of the basis polynomials as $V(\Phi) = V(p_a(\Phi))$. The potential has the same range as a function of the invariant polynomials, but is not affected by the degeneracy under group transformations. Since the polynomial basis is invariant under gauge transformations, a gauge orbit corresponds to a single point in the orbit space. Finding the bounded-from-below conditions and minimising the potential in the orbit space $\mathbb{R}^{n}/G$, with $p_{a}$ parameterising the orbits, can hence be advantageous \cite{Abud:1981tf,Kim:1981xu,Abud:1983id,Kim:1983mc}.

Each subgroup of the full gauge group $G$ is the isotropy subgroup $G_{\Phi}$ of some field configuration $\Phi$, and all the transformed states $\Phi_{\theta}$ have the same isotropy subgroup. The set or sub-manifold of gauge orbits that respects the same isotropy subgroup is called the stratum of the isotropy subgroup. It can be described by a finite number of polynomial equations or inequalities. The main stratum corresponds to a general field configuration and breaks the gauge symmetry completely. The field configurations that breaks the full gauge group $G$ to $G_{\Phi}$ lie in the stratum of $G_{\Phi}$. Thus the main stratum forms the interior of the orbit space, whereas the lower-dimensional strata that form the orbit space boundary correspond to more symmetrical field configurations invariant under different isotropy subgroups $G_{\Phi}$. The physical region of the invariant polynomials $p_{a}$ is called the orbit space. Therefore the orbit space consists of strata of different dimensions: vertices, edges, faces, \ldots, up to the main stratum whose dimension is given by the number of invariants.

The boundary of the orbit space can be found with the help of the $P$-matrix method \cite{Abud:1981tf,Abud:1983id,Talamini:2006wd}. The equations and inequalities that define the strata can be expressed in the form of positivity and rank conditions of a matrix $P$ constructed from the gradients of the polynomial invariants $p_{a}$. The $P$-matrix is the Hermitian square of the Jacobian matrix of gauge invariants, a $q \times q$ symmetric and positive semi-definite matrix, whose elements are constructed from gradients of basis invariants $p_{a}$ as
\begin{equation}
  P_{ab} = \frac{\partial p_{a}}{\partial \Phi_{i}^{*}} \frac{\partial p_{b}}{\partial \Phi_{i}},
\label{eq:p:matrix:def}
\end{equation}
where $\Phi_{i}$ run over the field components (real or complex). It can be shown that the $P$-matrix elements are invariant under gauge transformations and can be given in terms of the minimal polynomial basis. The $P$-matrix is positive-definite only inside the physical orbit space. The boundary of the orbit space can be obtained by solving the polynomial equation $\det P = 0$. The $k$-dimensional strata of the orbit space are given by the connected components of the set $p_{a}$ for which $P$ is positive-semidefinite and $\operatorname{rank} P = k$. For example, the full orbit space has a single vertex where the norms of fields vanish, given by vanishing order-1 minors of the $P$ matrix.

The gauge invariants can be separated into the norms of fields and angular variables or orbit space variables. Henceforth, by orbit space we will mostly mean the reduced orbit space of these dimensionless variables. For the reduced orbit space of the dimensionless orbit variables, vertices are given by the minors at the lowest order at which the field norms do not vanish, edges at the next order, and so on. If the scalar potential depends on the orbit space variables linearly, then its minimum lies on the boundary of the orbit space (more precisely, on the intersection of the orbit space with its convex hull) \cite{Kim:1981xu,Degee:2012sk,Heikinheimo:2017nth}. One does not have to minimise the potential over any flat or concave regions of the orbit space, since such a region is already accounted for in the convex hull by the boundary of the region.

The minimal integrity basis of polynomial invariants can involve higher-order invariants with dimension $d > 4$. It is straightforward to find such a basis. If the $P$-matrix for given invariants contains elements that cannot be expressed solely in terms of these invariants, we can extend the basis by these $P$-matrix elements and calculate a new $P$-matrix for the extended basis. The procedure can be repeated until all $P$-matrix elements can be expressed via the basis. It can be, however, very complicated to solve for the boundary of the orbit space and project out the higher-order invariants not present in the scalar potential. A more practical solution is to consider a partial $P$-matrix only for a partial basis of the invariants that appear in the scalar potential (a principal submatrix of the full $P$-matrix). The polynomial equation $\det P = 0$ can then be solved directly in terms of field components. Because the orbit space boundary corresponds to more symmetric field configurations, such solutions can often be found by educated guesswork.


\subsection{Orbit space for the Higgs doublet and a quadruplet}
\label{sec:orbit:space:P:H:phi}

In order to ascertain the boundary of the orbit space, we need to calculate the $P$-matrix and solve the equation $\det P = 0$. We calculate the $P$-matrix for the polynomial invariants present in the potential, namely the partial basis
\begin{equation}
\begin{aligned}
  p_{1} &= H^{*i} H_{i},
  &
  p_{2} &= \phi^{*ijk} \phi_{ijk},
  \\
  p_{3} &= \phi^{*ijk} \phi_{ijn} \phi^{*lmn} \phi_{lmk},
  &
  p_{4} &= H^{*i} H_{k} \phi^{*ljk} \phi_{lji},
  \\
  p_{5} &= \frac{1}{2} (\phi^{*ijk} H_{i} H_{j} \epsilon_{kl} H^{*l} + H_{l} \epsilon^{lk} H^{*i} H^{*j} \phi_{ijk})
  & \text{or \quad} 
  & \frac{1}{2} (\phi^{*ijk} H_{i} H_{j} H_{k} + \phi_{ijk} H^{*i} H^{*j} H^{*k}),
\end{aligned}
\label{eq:invariants}
\end{equation}
where the two definitions of $p_{5}$ are given for the $Y = 1/2$ and the $Y = 3/2$ quadruplet, respectively.\footnote{We choose to include a factor of $1/2$ in the definition of the $p_5$ invariant to make easier comparison with the type II seesaw and to avoid factors of $1/2$  appearing instead in the description of the shape of the orbit space.} Which one we use shall be clear from the context. 
We define, in terms of the gauge invariants in the potential, the dimensionless orbit space variables
\begin{equation}
  \zeta = \frac{p_{3}}{p_{2}^{2}},
  \qquad
  \xi = \frac{p_{4}}{p_{1} p_{2}},
  \qquad
  \chi = \frac{p_{5}}{\sqrt{p_{1}^{3} p_{2}}}.
\label{eq:orbit:space:variables:def}
\end{equation}
 The main stratum of the (reduced) orbit space is thus three-dimensional and the boundary of the orbit space has two-dimensional faces whose borders are edges that end at the vertices of the orbit space.
The field norms are given by
\begin{equation}
  \frac{h^{2}}{2} = p_{1}, \qquad \frac{\varphi^{2}}{2} = p_{2},
\end{equation}
where a coefficient of $1/2$ is included in the definition of $\varphi$ in order to simplify the potential. While the full orbit space contains both the orbit space variables and field norms, usually by orbit space we will mean the space of the dimensionless variables \eqref{eq:orbit:space:variables:def}.

The scalar potential, in terms of orbit space variables and the field norms, is then given by
\begin{equation}
\begin{split}
  V &= \frac{1}{2} \mu_{H}^{2} h^{2} + \frac{1}{2} \mu_{\phi}^{2} \varphi^{2} 
  + \frac{1}{4} \lambda_{H} h^{4} + \frac{1}{4} \lambda_{\phi} \varphi^{4} 
  + \frac{1}{4} \lambda'_{\phi} \zeta \varphi^{4} + \frac{1}{4} \lambda_{H\phi} h^{2} \varphi^{2} 
  \\
  & + \frac{1}{4} \lambda'_{H\phi} \xi h^{2} \varphi^{2} + \frac{1}{2} \hat{\lambda}_{H\phi} \chi h^{3} \varphi.
\end{split}
\label{eq:V:orbit}
\end{equation}
Notice that the potential looks the same, in terms of the orbit space variables, for both $Y = 1/2$ and $Y = 3/2$ quadruplets.

We can easily obtain partial knowledge about the orbit space. Defining a $2 \times 2$ Hermitian matrix $M^{k}_{l} \equiv \phi^{*ijk} \phi_{ijl}$, we see that $p_{2} = \tr M$, $p_{3} = \tr M^{2}$ and $p_{4} = H^{\dagger} M H$. We then have 
\begin{equation}
  \zeta = \frac{\tr M^{2}}{(\tr M)^{2}}, \qquad \xi = \frac{H^{\dagger} M H}{H^{\dagger} H \tr M}.
\end{equation}
In the basis with diagonal $M$, one can see that $\frac{1}{2} \leq \zeta \leq 1$ by considering equal diagonal elements or setting one diagonal element zero. The Cauchy-Schwartz inequality implies $0 \leq \xi \leq 1$. Considering also unit norms $H^\dagger H = 1$, $\tr M = 1$, one can readily show that the boundary of the projection of the orbit space to the $\zeta\xi$-plane is given by the $\zeta = 1 - 2 \xi +2 \xi^{2}$ curve.\footnote{Note that this dependence is exactly the same as for the corresponding orbit space variables of the type II seesaw scalar potential with the Higgs doublet and an $SU(2)_{L}$ triplet $\Delta$ \cite{Bonilla:2015eha}. This is not a coincidence, since the triplet is also presented by a $2 \times 2$ matrix, albeit a  traceless one. In fact, the orbit space of the $Y = 3/2$ quadruplet and the type II seesaw (with the type II seesaw $\chi$ defined for the $H^{T} (i \sigma_{2}) \Delta^{\dagger} H$ term) coincide \cite{KannikeInPrep}.} Furthermore, it is easy to see that $-1 \leq \chi \leq 1$ for the $Y = 3/2$ quadruplet. 

To fully determine the orbit space, we need to calculate the $P$-matrix.  
The symmetric $P$-matrix for the gauge invariants with $d \leq 4$ is given by
\begin{equation}
  P = 
  \begin{pmatrix}
    2 p_{1} & 0 & 0 & 2 p_{4} & 3 p_{5}
    \\
    \cdot & 2 p_{2}  & 4 p_{3} & 2 p_{4} & p_{5}
    \\
    \cdot & \cdot & P_{33} & P_{34} & P_{35}
    \\
    \cdot & \cdot & \cdot & P_{44} & P_{45}
    \\
    \cdot & \cdot & \cdot & \cdot & P_{55}
  \end{pmatrix},
\label{eq:p:matrix}
\end{equation}
where in the case of the $Y = 3/2$ quadruplet, we obtain
\begin{equation}
\begin{split}
  P_{33} &= \frac{8}{3} (\phi^{* ijk} \phi_{jkl} \phi^{* lmn} \phi_{mnp} \phi^{* pqr} \phi_{qri}
  + 2 \phi^{* ijk} \phi_{jkl} \phi^{* lmn} \phi_{imp} \phi^{* pqr} \phi_{qrn}), 
  \\
  P_{34} &= \frac{2}{3} ( - p_{1} p_{2}^{2} + p_{1} p_{3} + 2 p_{2} p_{4} 
  + 4 \phi^{* ijk} \phi_{ijp} \phi^{* pnm} \phi_{klm} H^{* l} H_{n}),
  \\
  P_{35} &= 2 \Re (H^{*i} H^{*j} \phi_{ijk} \phi^{*klm} \phi_{lmn} H^{*n}),
  \\
  P_{44} &= \frac{2}{3} \left( p_{1} p_{4} + 2 H^{*i} H^{*j} \phi_{ijk} \phi^{*klm} H_{l} H_{m} \right)
  - p_{1} p_{2}^{2} + p_{1} p_{3} + 2 p_{2} p_{4},
  \\
  P_{45} &= p_{1} p_{5} 
  + 3 \Re (H^{*i} \phi_{ijk} \phi^{*jkl} \phi_{lmn} H^{*m} H^{*n}),
  \\
  P_{55} &= \frac{1}{2} p_{1}^{3} 
  + \frac{9}{2} H^{* i} H^{* j} \phi_{ijk} \phi^{* klm} H_{l} H_{m}.
\end{split}
\label{eq:p:matrix:longer:elems}
\end{equation}
We computed the $P$-matrix from Eq. \eqref{eq:p:matrix:def} with the Wolfram Mathematica computer algebra system, but the compact expressions in Eq.~\eqref{eq:p:matrix:longer:elems} were calculated in birdtrack notation \cite{Cvitanovic:1976am,Cvitanovic:2008zz} in Appendix \ref{sec:pmatrix}. The birdtrack calculation provides a cross-check for the code.

With the Higgs boson in the unitary gauge, the orbit space for the Higgs doublet and the $Y = 3/2$ quadruplet is given by
\begin{align}
  \xi &= \frac{1}{3} \, \frac{\abs{\phi_{112}}^{2} 
  + 2 \abs{\phi_{122}}^{2} + 3 \abs{\phi_{222}}^{2}}{\abs{\phi_{111}}^{2} + \abs{\phi_{112}}^{2} 
  + \abs{\phi_{122}}^{2} + \abs{\phi_{222}}^{2}},
\label{eq:xi:Y:3:2:expr}
  \\
  \zeta &= \frac{1}{9} \, \Big[(3 \abs{\phi_{111}}^{2} + 2 \abs{\phi_{112}}^{2} + \abs{\phi_{122}}^{2})^{2} 
  + 2 \abs{\sqrt{3} \phi_{112}^{*} \phi_{111} + 2 \phi_{112} \phi_{122}^{*} + \sqrt{3} \phi_{122} \phi_{222}^{*}}^{2}
  \notag
  \\
  &+ (\abs{\phi_{111}}^{2} + 2 \abs{\phi_{122}}^{2} + 3 \abs{\phi_{222}}^{2})^{2}\Big]
  \Big/\Big[(\abs{\phi_{111}}^{2} + \abs{\phi_{112}}^{2} 
  + \abs{\phi_{122}}^{2} + \abs{\phi_{222}}^{2})^{2} \Big],
\label{eq:zeta:Y:3:2:expr}
  \\
  \chi &= \frac{\Re \phi_{222}}{\sqrt{\abs{\phi_{111}}^{2} + \abs{\phi_{112}}^{2} 
  + \abs{\phi_{122}}^{2} + \abs{\phi_{222}}^{2}}}.
\label{eq:chi:Y:3:2:expr}
\end{align}
Note that the Higgs field has canceled out of the orbit space variables which thus depend on the quadruplet only. This is due to our choice of unitary gauge and homogeneity of the invariant polynomials in $h$. In the limit of $h \to 0$, l'H\^{o}pital's rule gives the same result.

The shape of the orbit space can now be found by considering the principal minors of the $P$-matrix. First-order principal minors give $p_{1} = p_{2} = 0$ (then, of course, all $p_{i} = 0$). Second-order principal minors yield, among other equations, $p_{1} p_{2} = 0$. Taking $p_{2} = 0$ implies $p_{1} = 0$. Taking $p_{1} = 0$ or $h = 0$ leaves a single equation which is not identically satisfied. Simplifying matters further, we consider the field configuration $\phi_{112} = \phi_{122} = 0$ invariant under the $U(1)$ subgroup of $SU(2)$, and consider only real fields. We obtain
\begin{equation}
  \phi_{111}^{3} \phi_{222} = \phi_{111} \phi_{222}^{3}.
\label{eq:vertex:Y:3:2}
\end{equation}
Inserting its solutions in Eqs. \eqref{eq:zeta:Y:3:2:expr}, \eqref{eq:xi:Y:3:2:expr} and \eqref{eq:chi:Y:3:2:expr}, we obtain, for the orbit variables $(\xi,\zeta,\chi)$, the solutions $(1,1,\pm 1)$, $(0,1,0)$, $(1/2, 1/2, \pm 1/\sqrt{2})$. These are vertices of the reduced orbit space \eqref{eq:orbit:space:variables:def} with nonzero field norms.

We next look at solutions with $\operatorname{rank} P = 3$ with vanishing order-3 principal minors. We again consider only real fields and set $\phi_{112} = \phi_{122} = 0$. Going to unit $p_{2} = 1$, we solve it e.g. for $\phi_{222}$. The orbit space is then a function of $\phi_{111}$ only. The solutions then give either $\phi_{111} = 0$ or $h = 0$. The former choice gives again the above vertex solutions, while inserting the latter one in Eqs. \eqref{eq:zeta:Y:3:2:expr}, \eqref{eq:xi:Y:3:2:expr} and \eqref{eq:chi:Y:3:2:expr} and eliminating $\phi_{111}$ from \eqref{eq:orbit:space:variables:def}, we see that the orbit space has an edge given by
\begin{equation}
  \xi = \chi^{2}, 
  \quad
  \zeta = 1 - 2 \xi + 2 \xi^{2},
  \quad
  -1 \leq \chi \leq 1,
\label{eq:edge:Y:3:2}
\end{equation}
bounded by vertices given by $\xi = 1$, $\zeta = 1$, $\chi = \pm 1$ which correspond to neutral field configurations. The other solutions to Eq.~\eqref{eq:vertex:Y:3:2}, listed above, in fact lie on the edge \eqref{eq:edge:Y:3:2}. We can also check that the order-3 and higher principal minors indeed do vanish for the field configuration $\phi_{112} = \phi_{122} = 0$. This means that the faces bounded by the edge are determined by the edge and do not bulge out. Therefore, the convex hull of the orbit space is determined by the edge \eqref{eq:edge:Y:3:2}.

Allowing for non-zero $\phi_{112}$ or $\phi_{122} = 0$ yields more solutions to the equation $\det P = 0$, but they stay within the convex hull of the solution \eqref{eq:edge:Y:3:2}. We also checked that the convex hull for points from a random scan is within the convex hull of the edge \eqref{eq:edge:Y:3:2}.

\subsection{Orbit space for the Higgs doublet and the $Y = 1/2$ quadruplet}
\label{sec:orbit:space:P:H:phi:1:2}

For the $Y = 1/2$ quadruplet, the derivation is similar and we will not provide it in as much detail. The orbit variables $\xi$ and $\zeta$ are given by Eqs. \eqref{eq:xi:Y:3:2:expr} and \eqref{eq:zeta:Y:3:2:expr}, respectively, while $\chi$, related to $p_{5}$, is given by
\begin{equation}
  \chi = \frac{1}{\sqrt{3}} \frac{\Re \phi_{122}}{\sqrt{\abs{\phi_{111}}^{2} + \abs{\phi_{112}}^{2} 
  + \abs{\phi_{122}}^{2} + \abs{\phi_{222}}^{2}}}.
\label{eq:chi:Y:1:2:expr}
\end{equation}
Similarly, only the fifth column (and row) of the $P$-matrix changes. As expected, $P_{15} = 3 p_{5}$ and $P_{25} = p_{5}$ just like in the $Y = 3/2$ case. The remaining elements are calculated in Appendix \ref{sec:pmatrix} and are given by
\begin{align}
  P_{35} &= -\frac{2}{3} \left( \Re \phi^{* ijk} \phi_{ijl} \phi_{kmn} H^{*m} H^{*n} \epsilon^{lp} H_{p} 
  + 2 \Re \phi^{* ijk} \phi_{ijl} \phi_{kmn} \epsilon^{np} H^{*m} H^{*l} H_{p} \right),
  \\
  P_{45} &= \frac{2}{3} p_{1} p_{5} - 2 \Re H^{*i} \phi_{ijk} \phi^{* jkl} \phi_{lmn} H^{*m}
  \epsilon^{np} H_{p}
  - \Re H^{*i} \phi_{ijk} \phi^{* jkl} \epsilon_{lm} \phi^{mnp}  H_{n} H_{p},
  \\
  P_{55} &= \frac{1}{6} p_{1}^{3} + 2 H^{*i} H_{l} \epsilon^{lj} \phi_{ijk} \phi^{* kmn} \epsilon_{np} H_{m}  H^{p} + 2 \Re H^{*i} H_{l} \epsilon^{lj} \phi_{ijk} \epsilon^{km} \phi_{mnp} H^{*n} H^{*p} \notag
  \\
  &+ \frac{1}{2} H_{i} H_{j} \phi^{*ijk} \phi_{klm} H^{*l} H^{*m}.
\end{align}

The orbit space has two edges which meet at vertices given by $\xi = 2/3$, $\zeta = 5/9$, $\chi = \pm 1/\sqrt{3}$ that correspond to real neutral field configurations. One of the edges, obtained by setting $\phi_{112} = \phi_{122} = 0$ and real field values, is similar to \eqref{eq:edge:Y:3:2}:
\begin{equation}
  \xi = 2 \chi^{2}, 
  \quad
  \zeta = 1 - 2 \xi + 2 \xi^{2},
  \quad
  -\frac{1}{\sqrt{3}} \leq \chi \leq \frac{1}{\sqrt{3}}.
\label{eq:edge:1:2:first}
\end{equation}
The other edge, obtained by setting $\phi_{111} = \phi_{112} = 0$, is
\begin{equation}
    \xi = 1 - \chi^{2}, 
  \quad
  \zeta = 1 - 4 \chi^{4},
  \quad
  -\frac{1}{\sqrt{3}} \leq \chi \leq \frac{1}{\sqrt{3}},
\label{eq:edge:1:2:last}
\end{equation}
and its projection on the $\xi\zeta$-plane is given by $\zeta = -3 + 8 \xi - 4 \xi^{2}$.
The orbit space is symmetric with respect to the $\chi = 0$ line, obtained by setting $\phi_{112} = \phi_{222} = 0$, and given by 
\begin{equation}
  0 \leq \xi \leq 1, 
  \quad
  \zeta = 1 - 2 \xi + 2 \xi^{2},
  \quad
  \chi = 0.
\label{eq:edge:1:2:central}
\end{equation}

\section{Bounded-from-below conditions}
\label{sec:bfb}

In order to make physical sense, the scalar potential of a theory must be bounded-from-below to have a finite potential minimum. Since dimensionful terms in the potential can be neglected for large field values, it is enough to only consider constraints on the quartic part of the potential. Because we can always take the $\hat{\lambda}_{H\phi}$ term to be negative without loss of generality, it suffices to consider non-negative $\chi$. 

First of all, simple sufficient bounded-from-below conditions can easily be obtained by taking $\lambda_{\phi} = \lambda'_{\phi} = 0$, because then the quartic part of the potential \eqref{eq:V:orbit} reduces to\footnote{Thanks to Gauthier Durieux, Matthew McCullough and Ennio Salvioni for pointing this out.}
\begin{equation}
  V = h^{2} \, [\lambda_{H} h^{2} + (\lambda_{H\phi} + \lambda'_{H\phi} \xi^{2}) \varphi^{2} - 2 |\hat{\lambda}_{H\phi}| \chi h \varphi].
\label{eq:V:orbit:no:self}
\end{equation}
Because the last term in brackets is always negative, we can consider usual positivity, not copositivity, of its coefficient matrix. 

\subsection{Bounded-from-below conditions for the $Y = 3/2$ quadruplet}
\label{sec:bfb:3:2}

Since the orbit space for the $Y = 3/2$ quadruplet is simpler, we will consider this case first. The discriminant of the potential \eqref{eq:V:orbit:no:self} with zero self-couplings is extremised for $\chi = 0$ or $\chi = 1$, yielding the sufficient bounded-from-below conditions
\begin{equation}
  \lambda_{H} > 0, 
  \quad 
  \lambda_{H\phi} > 0, 
  \quad 
  \lambda_{H\phi} + \lambda'_{H\phi} > 0,
  \quad
  \lambda_{H} (\lambda_{H\phi} + \lambda'_{H\phi}) > \hat{\lambda}_{H\phi}^{2}.
\label{eq:bfb:sufficient:3:2}
\end{equation}

For the full conditions, we have to minimise the quartic potential on the boundary of the orbit space, which in our case means its vertices and edges. Because we can consider only non-negative $\chi$, we have to consider also the middle of the orbit space curved edge, at $\chi = 0$, where we have $\zeta = 0$, $\xi = 1$. The potential is biquadratic in field norms for $\chi = 0$, so we can get the vacuum stability conditions from copositivity \cite{Kannike:2012pe} at this point. At the $\chi = 1$ vertex of the orbit space, we can consider positivity of the potential as a quartic polynomial in $\varphi$ with $h = 1$ \cite{Kannike:2016fmd}. We obtain the necessary vacuum stability conditions for the potential as
\begin{equation}
\begin{split}
  \lambda_{H} > 0, 
  \quad 
  \lambda_{\phi} > 0, 
  \quad 
  \lambda_{\phi} + \lambda'_{\phi} &> 0,
  \\
  \lambda_{H\phi} + 2 \sqrt{\lambda_{H} (\lambda_{\phi} + \lambda'_{\phi})} > 0,
  \quad 
  \lambda_{H\phi} + \lambda'_{H\phi} + 2 \sqrt{\lambda_{H} (\lambda_{\phi} + \lambda'_{\phi})} &> 0, 
  \\
  (\lambda_{\phi} + \lambda'_{\phi}) \Big( \lambda_{H} [(\lambda_{H\phi} + \lambda'_{H\phi})^2 
  - 4 (\lambda_{\phi} + \lambda'_{\phi}) \lambda_{H}]^2 
  - (\lambda_{H\phi} + \lambda'_{H\phi}) \\
  \times [(\lambda_{H\phi} + \lambda'_{H\phi})^2 
  - 36 (\lambda_{\phi} + \lambda'_{\phi}) \lambda_{H}] \hat{\lambda}_{H\phi}^2 - 
 27  (\lambda_{\phi} + \lambda'_{\phi}) \hat{\lambda}_{H\phi}^4 \Big) &> 0.
  \end{split}
\label{eq:bfb:necessary:3:2}
\end{equation}
Note that the last, most complex condition in \eqref{eq:bfb:necessary:3:2} only depends on the combinations $\lambda_{\phi} + \lambda'_{\phi}$ and $\lambda_{H\phi} + \lambda'_{H\phi}$. For $h = 0$, we obtain further necessary conditions
\begin{equation}
  2\lambda_{\phi} + \lambda'_{\phi} > 0, \quad \lambda_{\phi} + \lambda'_{\phi} > 0,
\end{equation}
of which the last one coincides with one of the conditions of Eq.~\eqref{eq:bfb:necessary:3:2}.

The full conditions are obtained by considering vacuum stability conditions on the edge of the orbit space, where the potential on a unit circle, $V + \lambda (1 - h^{2} - \varphi^{2})$, can be minimised with respect to the parameter $\chi$ on the edge \eqref{eq:edge:Y:3:2}, the fields $h$ and $\varphi$, and the Lagrange multiplier $\lambda$ enforcing the constraint. The minimisation equations, given by
\begin{equation}
\begin{aligned}
  0 &= \varphi \, [ 2 \lambda'_{\phi} \chi (2 \chi^{2} - 1) \varphi^{3} + 
  \lambda'_{H\phi} \chi h^{2} \phi - \hat{\lambda}_{H\phi} h^{3}],
  \\
  \lambda h &= h[4 \lambda_{H} h^{2} + 2 (\lambda_{H\phi} + \lambda'_{H\phi} \chi^{2}) \varphi^{2}
  -6 \hat{\lambda}_{H\phi} \chi h \varphi],
  \\
  \lambda \varphi &= 4 [\lambda_{\phi} + \lambda'_{\phi} (1 - 2 \chi^{2} + 2 \chi^{4})] \varphi^{3}
  + 2 (\lambda_{H\phi} + \lambda'_{H\phi} \chi^{2}) h^{2} \varphi - 2 \hat{\lambda}_{H\phi} \chi h^{3},
  \\
  1 &= h^{2} + \varphi^{2},
\end{aligned}
\label{eq:V:min:unit:circle:edge:3:2}
\end{equation}
can only be solved numerically. The bounded-from-below conditions on the edge are given by the intersection of the condition
\begin{equation}
  0 < \chi < 1 \land 0 < h < 1 \land 0 < \varphi < 1 \implies V > 0
\label{eq:bfb:edge}
\end{equation}
for each real solution of the minimisation equations. The necessary and sufficient conditions are then given by Eqs. \eqref{eq:bfb:necessary:3:2} and \eqref{eq:bfb:edge}. In practice, it can be faster to consider the positivity of the potential as a quartic polynomial in $\varphi$ with $h = 1$ and $\chi$ on the edge for a range of $\chi$ values, including the end points $\chi = 0$ and $\chi = 1$. 

The parameter space boundary given by full necessary and sufficient vacuum stability conditions lies between the bounds given by Eq.~\eqref{eq:bfb:necessary:3:2} and Eq.~\eqref{eq:bfb:sufficient:3:2}. The difference is large only for when the quadruplet self-couplings are sizeable in comparison with its Higgs portal couplings. 

\subsection{Bounded-from-below conditions for the $Y = 1/2$ quadruplet}
\label{sec:bfb:1:2}

The sufficient conditions for the potential \eqref{eq:V:orbit:no:self} with zero self-couplings are given by
\begin{equation}
  \lambda_{H} > 0, 
  \quad 
  \lambda_{H\phi} > 0, 
  \quad 
  \lambda_{H\phi} + \lambda'_{H\phi} > 0,
  \quad
  \lambda_{H} (3 \lambda_{H\phi} + 2 \lambda'_{H\phi}) > \hat{\lambda}_{H\phi}^{2},
\label{eq:bfb:sufficient:1:2}
\end{equation}
where we considered positivity of the potential \eqref{eq:V:orbit:no:self} for the extremal values $\chi = 0$ and $\chi = 1/\sqrt{3}$. 

Considering copositivity at the end and middle points of the $\chi = 0$ line \eqref{eq:edge:1:2:central}, and positivity of the potential as a quartic polynomial in $\varphi$ with $h = 1$ for the $\chi = 1/\sqrt{3}$ vertex of the orbit space, we find the necessary conditions
\begin{equation}
\begin{split}
  \lambda_{H} > 0, 
  \quad 
  \lambda_{\phi} + \lambda'_{\phi} > 0, 
  \quad 
  2 \lambda_{\phi} + \lambda'_{\phi} > 0,
  \quad
  9 \lambda_{\phi} + 5 \lambda'_{\phi} &> 0,
  \\
  \lambda_{H\phi} + 2 \sqrt{\lambda_{H} (\lambda_{\phi} + \lambda'_{\phi})} > 0,
  \quad 
  \lambda_{H\phi} + \lambda'_{H\phi} 
  + 2 \sqrt{\lambda_{H} (\lambda_{\phi} + \lambda'_{\phi})} &> 0, 
  \\
  3 \lambda_{H\phi} + 2 \lambda'_{H\phi} 
  + 2 \sqrt{\lambda_{H} (9 \lambda_{\phi} + 5 \lambda'_{\phi})} &> 0,
  \\
  (9 \lambda_{\phi} + 5 \lambda'_{\phi}) 
  \Big( \lambda_{H} [(3 \lambda_{H\phi} - 2 \lambda'_{H\phi})^{2} - 4 \lambda_{H} (9 \lambda_{\phi} 
  + 5 \lambda'_{\phi})]^{2} &
  \\
  + [36 \lambda_{H} (9 \lambda_{\phi} + 5 \lambda'_{\phi}) (3 \lambda_{H\phi} + 2 \lambda'_{H\phi})
  - (3 \lambda_{H\phi} + 2 \lambda'_{H\phi})^{3}] \hat{\lambda}_{H\phi}^2
   - 27 (9 \lambda_{\phi} + 5 \lambda'_{\phi}) \hat{\lambda}_{H\phi}^4 \Big) &> 0.
  \end{split}
\label{eq:bfb:necessary:1:2}
\end{equation}

The bounded-from-below conditions on the edges \eqref{eq:edge:1:2:first} and \eqref{eq:edge:1:2:last} are given by the intersection of
\begin{equation}
  0 < \chi < 1/\sqrt{3} \land 0 < h < 1 \land 0 < \varphi < 1 \implies V > 0
\label{eq:bfb:edges:1:2}
\end{equation}
for real minimum solutions of the potential on a unit circle $V + \lambda (1 - h^{2} - \varphi^{2})$ with the orbit space parameters, respectively, on each edge. The minimum equations are similar to Eq. \eqref{eq:V:min:unit:circle:edge:3:2}. On the central line \eqref{eq:edge:1:2:central}, the potential is copositive and the bounded-from-below conditions have the same form as the conditions for the type II seesaw potential derived in Ref. \cite{Bonilla:2015eha}:
\begin{equation}
\begin{split}
  \lambda_{H} > 0, 
  \quad
  \lambda_{\phi} + \lambda'_{\phi} > 0, 
  \quad
  2 \lambda_{\phi} + \lambda'_{\phi} &> 0,
  \\
  \lambda_{H\phi} + 2 \sqrt{\lambda_{H} (\lambda_{\phi} + \lambda'_{\phi})} > 0,
  \quad 
  \lambda_{H\phi} + \lambda'_{H\phi} 
  + 2 \sqrt{\lambda_{H} (\lambda_{\phi} + \lambda'_{\phi})} &> 0, 
  \\
  \sqrt{\lambda_{\phi} + \lambda'_{\phi}} \abs{\lambda'_{H\phi}} 
  \geq 2 \sqrt{\lambda_{H}} \lambda'_{\phi} 
  \quad \lor \quad 
  \lambda_{H\phi} + \frac{1}{2} \lambda'_{H\phi} 
  + \frac{1}{2} \sqrt{\left(2 \frac{\lambda_{\phi}}{\lambda'_{\phi}} + 1 \right)
  \left(8 \lambda_{H} \lambda'_{\phi} - \lambda_{H\phi}^{\prime 2} \right)} &>0.
\end{split}
\label{eq:bfb:edge:central:1:2}
\end{equation}
The full necessary and sufficient bounded-from-below conditions are given by Eqs. \eqref{eq:bfb:necessary:1:2}, \eqref{eq:bfb:edges:1:2} and \eqref{eq:bfb:edge:central:1:2}. Just as for the $Y = 3/2$ case, the parameter space boundary given by full necessary and sufficient vacuum stability conditions lies between the bounds given by Eq.~\eqref{eq:bfb:sufficient:1:2} and Eq.~\eqref{eq:bfb:necessary:1:2}. The difference is large only for when the quadruplet self-couplings are sizeable in comparison with its Higgs portal couplings. 

\section{Perturbative unitarity}
\label{sec:pert:unit}

First of all, we require each quartic coupling to be less than $4 \pi$ in absolute value in order for the theory to be perturbative. Further, perturbative unitarity constraints arise from the unitarity of the scattering matrix for the two-to-two scalar scattering amplitudes. At the order of the zeroth partial wave, the $S$-matrix is given by
\begin{equation}
  a_{0}^{ba} = \sqrt{\frac{4 \abs{\mathbf{p}_{b}} \abs{\mathbf{p}_{a}}}{2^{\delta_{a}} 2^{\delta_{b}} s}}
  \int_{-1}^{1} d(\cos \theta) \mathcal{M}_{ba}(\cos \theta)
\end{equation}
with a pair $a$ of scalars scattering to another pair $b$ with the matrix element $\mathcal{M}_{ba}(\cos \theta)$. The angle $\theta$ is between the incoming ($\mathbf{p}_{a}$) and outgoing ($\mathbf{p}_{b}$) three-momenta in the centre-of-mass frame, and the Mandelstam variable $s = (p_{1} + p_{2})^{2}$. The exponent $\delta_{a}$  is unity if the particles in pair $a$ are identical and zero otherwise; likewise for pair $b$ and $\delta_{b}$. The eigenvalues $a_{0}^{i}$ of the scattering matrix must satisfy
\begin{equation}
  \abs{\Re a_{0}^{i}} \leq \frac{1}{2}.
\end{equation}

In the high energy limit with infinite scattering energy $s \to \infty$ only quartic couplings contribute to scattering. We calculate the transition matrix for all possible two-to-two scatterings $S_{1} S_{2} \to S_{3} S_{4}$ of scalar bosons, including the Goldstones $G^{0}$ and $G^{+}$. In this limit, the $a_{0}^{ba}$ matrix element, where $a \equiv S_{1} S_{2}$ and $b \equiv S_{3} S_{4}$, is given by 
\begin{equation}
  a_{0}^{ba} = \frac{1}{16 \pi} \frac{1}{\sqrt{2^{\delta_{S_{1} S_{2}}} 2^{\delta_{S_{3} S_{4}}}}} \frac{\partial^{4} V}{\partial S_{1} \partial S_{2} \partial S_{3}^{*} \partial S_{4}^{*}}.
\end{equation}
We do not give an explicit list of possible initial and final states for the $Y = 1/2$ and $Y = 3/2$ cases nor the $S$-matrices, because they are very lengthy and it is straightforward to obtain them.

\begin{figure}[p]
\begin{center}
  \includegraphics{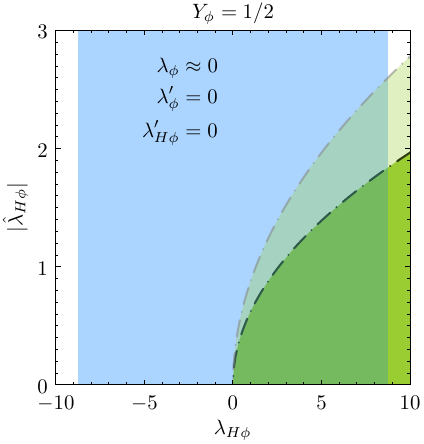}
  \quad
  \includegraphics{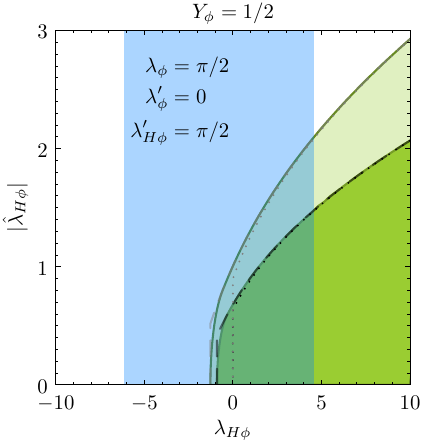}
  \\
  \vspace{1em}
  \includegraphics{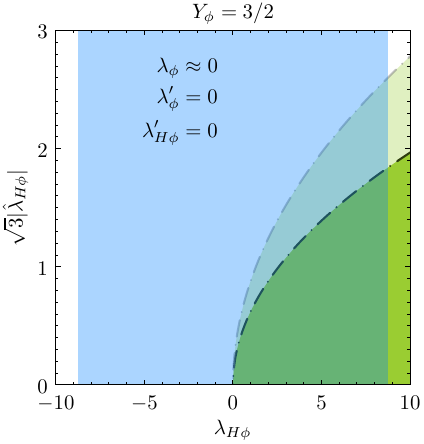}
  \quad
  \includegraphics{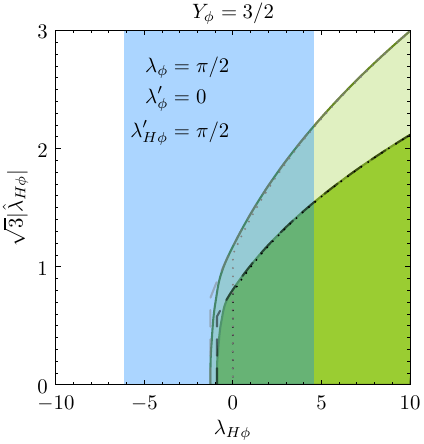}
\caption{Parameter space for the $Y = 1/2$ and $Y = 3/2$ quadruplets allowed by bounded-from-below constraints, where green stands for $\lambda_{H} = \lambda_{H}^{\rm SM} = 0.129$ and light green for $\lambda_{H} = 2 \lambda_{H}^{\rm SM}$, together with the boundaries of the respective sufficient (dotted) and necessary (dashed) bounded-from-below constraint regions. Perturbative unitarity, which changes very little between the two  values of $\lambda_{H}$, is satisfied in the blue region. For $Y = 3/2$ we show $\sqrt{3} |\hat{\lambda}_{H\phi}|$ on the vertical axis to facilitate comparison with Ref.~\cite{Durieux:2022hbu}.}
\label{fig:bfb:quadruplet}
\end{center}
\end{figure}

Most unitarity constraints, however, cannot be expressed analytically. Still, we can considerably simplify the unitarity constraints by considering them together with other conditions. We see that in practice the bounded-from-below conditions on the $\hat{\lambda}_{H\phi}$ coupling are much stronger than unitarity constraints which therefore practically do not depend on this coupling in the allowed region. In this approximation, the perturbative unitarity conditions are given by
\begin{equation}
\begin{aligned}
    \abs{\lambda_{H}} &< 2 \pi, &
    \abs{\lambda_{\phi} + \lambda'_{\phi}} &< 2 \pi, &
    \abs{9 \lambda_{\phi} - \lambda'_{\phi}} &< 36 \pi, &
    \\
    \abs{9 \lambda_{\phi} + 5 \lambda'_{\phi}} &< 18 \pi, &
    \abs{3 \lambda_{H\phi} - \lambda'_{H\phi}} &< 24 \pi, &
    \abs{3 \lambda_{H\phi} + 4 \lambda'_{H\phi}} &< 24 \pi,
    \\
    \span \span \span \span
    \abs{9 \lambda_{H} + 15 \lambda_{\phi} + 10 \lambda'_{\phi} 
    + \sqrt{(15 \lambda_{\phi} + 10 \lambda'_{\phi} -9 \lambda_{H})^{2} 
    + 18 (2 \lambda_{H\phi} + \lambda'_{H\phi})^{2}}} &< 24 \pi,
    \\
    \span \span \span \span
    \abs{9 \lambda_{H} + 15 \lambda_{\phi} + 10 \lambda'_{\phi} 
    - \sqrt{(15 \lambda_{\phi} + 10 \lambda'_{\phi} -9 \lambda_{H})^{2} 
    + 18 (2 \lambda_{H\phi} + \lambda'_{H\phi})^{2}}} &< 24 \pi,
    \\
    \span \span \span \span
    \abs{9 \lambda_{H} + 9 \lambda_{\phi} + 20 \lambda'_{\phi} 
    + \sqrt{(9 \lambda_{\phi} + 20 \lambda'_{\phi} -9 \lambda_{H})^{2} 
    + 90 \lambda_{H\phi}^{\prime 2}}} &< 24 \pi,
    \\
    \span \span \span \span
    \abs{9 \lambda_{H} + 9 \lambda_{\phi} + 20 \lambda'_{\phi} 
    - \sqrt{(9 \lambda_{\phi} + 20 \lambda'_{\phi} -9 \lambda_{H})^{2} 
    + 90 \lambda_{H\phi}^{\prime 2}}} &< 24 \pi,
\end{aligned}
\label{eq:pert:unit:approx}
\end{equation}
where we have omitted conditions that are already guaranteed by perturbativity.
Note that because we set $\hat{\lambda}_{H\phi} = 0$, the approximate constraints \eqref{eq:pert:unit:approx} apply both in the $Y = 1/2$ and $Y = 3/2$ cases. On the other hand, we can set all couplings save $\hat{\lambda}_{H\phi}$ to zero to get an upper bound, finding that
\begin{equation}
  \abs{\hat{\lambda}_{H\phi}} < 2 \sqrt{6} \pi \text{ for $Y = \frac{1}{2}$ \quad and \quad }
  \abs{\hat{\lambda}_{H\phi}} < \sqrt{\frac{2}{3}} 4 \pi \text{ for $Y = \frac{3}{2}$}.
\end{equation}

\section{Results}
\label{sec:results}

Figure~\ref{fig:bfb:quadruplet} shows examples of the parameter space for the $Y = 1/2$ (top row) and $Y = 3/2$ (bottom row) quadruplets allowed by bounded-from-below conditions and perturbative unitarity constraints \eqref{eq:pert:unit:approx}. Green colour stands for $\lambda_{H} = \lambda_{H}^{\rm SM} = 0.129$, and light green for $\lambda_{H} = 2 \lambda_{H}^{\rm SM}$. Also shown are the boundaries of the respective sufficient (dotted) and necessary (dashed) bounded-from-below constraint regions. The left column of Figure~\ref{fig:bfb:quadruplet} shows the allowed region for $\lambda_{\phi} \approx 0$ (that is, $0 < \lambda_{\phi} \ll 1$), $\lambda'_{\phi} = 0$, $\lambda'_{H\phi} = 0$, and the right column for $\lambda_{\phi} = \pi/2$, $\lambda'_{\phi} = 0$, $\lambda'_{H\phi} = \pi/2$. Regions allowed by  unitarity constraints (blue) practically do not change between the two values of $\lambda_{H}$. 

Note that the $\mathbb{Z}_{2}$-breaking term \eqref{eq:V:Z:2:breaking:Y:1:2} for the $Y = 1/2$ quadruplet has the same normalisation than Ref.~\cite{Durieux:2022hbu}, while for the $Y = 3/2$ quadruplet we have $\hat{\lambda}_{H\phi} = \lambda_{\tilde{\Phi}}/\sqrt{3}$. For this reason we display the bounds for $\sqrt{3} \hat{\lambda}_{H\phi}$ for the $Y = 3/2$ quadruplet so as to be able to directly compare results. We see that the bounds from our conditions on $\lambda_{\Phi}$ and $\lambda_{\tilde{\Phi}}$ of Ref.~\cite{Durieux:2022hbu} are roughly the same. Note that for very small quadruplet self-couplings it suffices to use the sufficient conditions \eqref{eq:bfb:sufficient:1:2} or \eqref{eq:bfb:sufficient:3:2}. But even finite self-couplings play a role only for small Higgs portals, in which case the necessary conditions \eqref{eq:bfb:necessary:3:2} are an excellent approximation to the full necessary and sufficient ones.

While larger self-couplings or Higgs portals do increase the region allowed by the bounded-from-below conditions, the corresponding decrease in the region allowed by perturbative unitarity constraints is more significant and undoes any gain. In case of the $Y = 1/2$ quadruplet, the combination of bounded-from-below conditions and unitarity gives an upper bound of $\hat{\lambda}_{H\Phi} \approx 1.8$ for $\lambda_{H} = \lambda_{H}^{\rm SM}$ and $\hat{\lambda}_{H\Phi} \approx 2.8$ for $\lambda_{H} = 2 \lambda_{H}^{\rm SM}$ (the same bounds apply to $\sqrt{3} \hat{\lambda}_{H\Phi}$ for the $Y = 3/2$ quadruplet). These bounds become stronger than the electroweak precision measurement bounds on models that violate custodial symmetry, elaborated on in Ref.~\cite{Durieux:2022hbu}, at quadruplet masses above about $M_{\phi} \approx 1.8$~TeV and at $M_{\phi} \approx 2.4$~TeV, respectively. 

The $\lambda_{H} = \lambda_{H}^{\rm SM}$ case then gives $- 0.16 \leq \delta_{h^{3}} \leq 0$ and the $\lambda_{H} = 2 \lambda_{H}^{\rm SM}$ gives $- 0.38 \leq \delta_{h^{3}} \leq 0$ for either the $Y = 1/2$ or $Y = 3/2$ quadruplet model. For the custodial quadruplet, which is not restricted by electroweak precision measurements, the bounds are 
\begin{equation}
  \frac{-1.0}{M_{\phi}/\mathrm{TeV}} \leq \delta_{h^{3}} \leq 0 
  \text{ for $\lambda_{H} = \lambda_{H}^{\rm SM}$} 
  \quad
  \text{and}
  \quad
  \frac{-2.5}{M_{\phi}/\mathrm{TeV}} \leq \delta_{h^{3}} \leq 0 
  \text{ for $\lambda_{H} = 2 \lambda_{H}^{\rm SM}$}.
\end{equation}

\section{Conclusions}
\label{sec:conclusions}

We have derived bounded-from-below conditions and perturbative unitarity constraints \eqref{eq:pert:unit:approx} of the Higgs doublet $H$ and an $SU(2)$ quadruplet $\phi$ of hypercharge $Y = 1/2$ or $Y = 3/2$. The new conditions constrain the $\hat{\lambda}_{H\phi}$ coupling that induces a change in the trilinear Higgs coupling upon integrating out the quadruplet. The available parameter space is shown in figure \ref{fig:bfb:quadruplet}. Smaller quadruplet self-couplings allow for larger values of $\hat{\lambda}_{H\phi}$, for though the bounded-from-below conditions are relaxed for larger self-couplings, the perturbative unitarity constraints become severe much faster. 

Although the $\hat{\lambda}_{H\phi}$ quartic is constrained by electroweak precision measurements for models that violate custodial symmetry, these constraints weaken with higher quadruplet masses. By comparison with the results of  Ref.~\cite{Durieux:2022hbu}, we show that the combination of bounded-from-below conditions and unitarity becomes stronger than electroweak precision constraints for quadruplet masses above a couple of TeV in such models. For the custodial quadruplet, which is not restricted by electroweak precision measurements, the bounded-from-below conditions in combination with unitarity can be the strongest constraint of all.

Calculating the bounded-from-below conditions is simplified by using the gauge orbit space whose shape is determined with the $P$-matrix formalism. We compute $P$-matrices via birdtrack formalism, given in Appendix \ref{sec:pmatrix}, that avoids keeping track of a multitude of explicit $SU(2)$ indices.

\appendix

\section{$P$-matrix for the Higgs doublet and a quadruplet}
\label{sec:pmatrix}

\subsection{Case of the $Y = 3/2$ quadruplet}
\label{sec:pmatrix:y:3:2}

The $P$-matrix \eqref{eq:p:matrix:def}, which determines the shape of the orbit space, can be calculated in a computer algebra system such as Wolfram Mathematica. Explicit calculations with field tensors are inconvenient with pen and paper since there is a large number of indices to be tracked. The $P$-matrix elements, however, can be calculated with birdtrack diagrams \cite{Cvitanovic:1976am,Cvitanovic:2008zz} (see \cite{Keppeler:2017kwt,Peigne:2023iwm} for pedagogical reviews). Because the case of the $Y = 3/2$ quadruplet is less complicated, we will deal with it first.

Let us denote the fields as
\begin{equation}
  H_{i} = \TextVCenter{\includegraphics{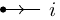}}, 
  \quad
  H^{* i} = \TextVCenter{\includegraphics{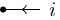}},
  \quad
  \phi_{ijk} = \TextVCenter{\includegraphics{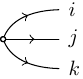}},
  \quad
  \phi^{*ijk} = \TextVCenter{\includegraphics{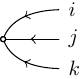}}.
\label{eq:fields:birdtracks}
\end{equation}
In the most part we will not attach names for $SU(2)$ indices to the legs.
Note that $\phi$ is a symmetric tensor,
\begin{equation}
  \phi_{ijk} = \phi_{(ijk)} \quad \text{or} \quad 
  \TextVCenter{\includegraphics{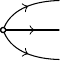}} \; = \; \TextVCenter{\includegraphics{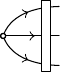}},
  \label{eq:phi:birdtracks}
\end{equation}
where the symmetriser for three lines is given by all their possible permutations:
\begin{equation}
  \TextVCenter{\includegraphics{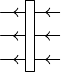}} = \frac{1}{3!} 
  \left( \, \TextVCenter{\includegraphics{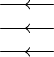}} \, + \, \TextVCenter{\includegraphics{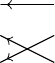}} 
  \, + \, \TextVCenter{\includegraphics{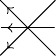}} \, + \, \TextVCenter{\includegraphics{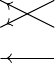}}
  \, + \, \TextVCenter{\includegraphics{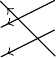}} \, + \, \TextVCenter{\includegraphics{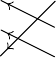}}
  \,  \right).
\label{eq:symm}
\end{equation}
Placing  diagrams next to each other denotes their product. In order to obtain the Hermitian conjugate of a diagram, we take its mirror image and reverse all arrows \cite{Peigne:2023iwm}. Our definitions ensure that we can rotate a diagram without changing its meaning. 

The gauge invariant polynomials \eqref{eq:invariants} are given by
\begin{equation}
\begin{split}
  p_{1} &= H^{*i} H_{i} = \TextVCenter{\includegraphics{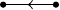}},
  \\
  p_{2} &= \phi^{*ijk} \phi_{ijk} = \TextVCenter{\includegraphics{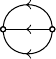}},
  \\
  p_{3} &= \phi^{*ijk} \phi_{ijn} \phi^{*lmn} \phi_{lmk} = \TextVCenter{\includegraphics{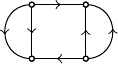}},
  \\
  p_{4} &= H^{*i} H_{k} \phi^{*ljk} \phi_{lji} = \TextVCenter{\includegraphics{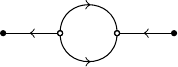}},
  \\
  p_{5} &= \Re (\phi^{*ijk} H_{i} H_{j} H_{k}) = \Re \,	 \TextVCenter{\includegraphics{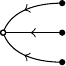}} =  \frac{1}{2} \left( \TextVCenter{\includegraphics{p5}} + \TextVCenter{\includegraphics{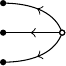}} \right),
\end{split}
\end{equation}
where a line from tensor to tensor means summation over the corresponding index.
We can write all scalar fields \eqref{eq:fields:birdtracks} in a single vector as
\begin{equation}
  \Phi = (\phi_{ijk}, \phi^{* lmn}, H_{p}, H^{*q}) =  
  \Big( \TextVCenter{\includegraphics{phi}}, \;
  \TextVCenter{\includegraphics{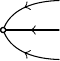}}, \;
  \TextVCenter{\includegraphics{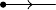}}, \;
  \TextVCenter{\includegraphics{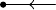}} \Big).
\label{eq:field:vector}
\end{equation}
Taking the derivative with respect to a scalar field corresponds to removing the corresponding field symbol, observing Leibniz's rule, from all possible places in the diagram one at the time. Therefore, the gradients of the gauge invariants with respect to the field vector \eqref{eq:field:vector} are given by
\begin{align}
  \frac{\partial{p_{1}}}{\partial \Phi} &=
  \Big(0, 0, \; \TextVCenter{\includegraphics{Hdagger}}, \; \TextVCenter{\includegraphics{H}} \Big),
  \label{eq:p1:grad}
  \\
  \frac{\partial{p_{2}}}{\partial \Phi} &=
  \Big(\; \TextVCenter{\includegraphics{phidagger}}, \; \TextVCenter{\includegraphics{phi}}, 0, 0 \Big),
  \label{eq:p2:grad}
  \\
  \frac{\partial{p_{3}}}{\partial \Phi} &=
  \left(\TextVCenter{\includegraphics{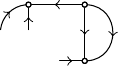}} + \TextVCenter{\includegraphics{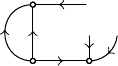}},
  \;
  \TextVCenter{\includegraphics{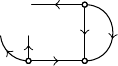}} + \TextVCenter{\includegraphics{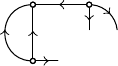}}, 0, 0 \right)
  \notag
  \\
  &= \left( 2 \, \TextVCenter{\includegraphics{p3grad11}}, \; 2 \, \TextVCenter{\includegraphics{p3grad21}}, 0, 0 \right),
  \\
  \frac{\partial{p_{4}}}{\partial \Phi} &= \left( \TextVCenter{\includegraphics{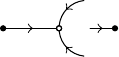}}, 
  \;
  \TextVCenter{\includegraphics{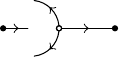}},
  \;
  \TextVCenter{\includegraphics{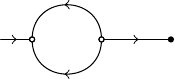}}, \;
  \TextVCenter{\includegraphics{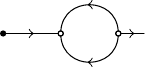}} \right),
  \\
  \frac{\partial{p_{5}}}{\partial \Phi} &= \frac{1}{2}
  \left(\TextVCenter{\includegraphics{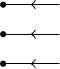}}, \; 
  \TextVCenter{\includegraphics{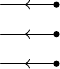}}, \;
  3 \, \TextVCenter{\includegraphics{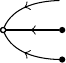}}, \;
  3 \, \TextVCenter{\includegraphics{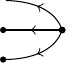}} \right).
\end{align}

We can now calculate the elements of the $P$-matrix. Note that when one of the gradients is of a field norm, i.e. $\partial p_{1}/\partial \Phi$ or $\partial p_{2}/\partial \Phi$, the answer is proportional to the other invariant:
\begin{align}
  P_{11} &= 2 \, \TextVCenter{\includegraphics{p1}} = 2 p_{1},
  \\
  P_{12} &= 0,
  \\
  P_{13} &= 0,
  \\
  P_{14} &= 2 \, \TextVCenter{\includegraphics{p4}} = 2 p_{4},
  \\
  P_{15} &= \frac{3}{2} \left( \TextVCenter{\includegraphics{p5}} + \TextVCenter{\includegraphics{p5conj}} \right) = 3 p_{5},
  \\
  P_{22} &= 2 \, \TextVCenter{\includegraphics{p2}} = 2 p_{2},
  \\
  P_{23} &= 4 \, \TextVCenter{\includegraphics{p3}} = 4 p_{3},
  \\
  P_{24} &= 2 \, \TextVCenter{\includegraphics{p4}} = 2 p_{4},
  \\
  P_{25} &= \frac{1}{2} \left( \TextVCenter{\includegraphics{p5}} + \TextVCenter{\includegraphics{p5conj}} \right) = p_{5}.
\end{align}
In some of the $P$-matrix elements there is no choice as to which lines to join. In calculating the $P_{33}$ and other similar elements, on the other hand, we have to keep in mind that there are several possible ways to join the lines, which we can take into account by inserting a symmetriser.\footnote{Note that this has nothing to do with $\phi$ being a symmetric tensor in Eq.~\eqref{eq:phi:birdtracks}.}
Therefore
\begin{equation}
\begin{split}
  P_{33} &= 8 \, \TextVCenter{\includegraphics{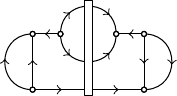}} 
  \\
  &= \frac{4}{3} \left(2 \TextVCenter{\includegraphics{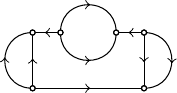}}
  + 4 \TextVCenter{\includegraphics{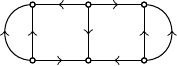}} \right).
\end{split}
\end{equation}
where we inserted the symmetriser \eqref{eq:symm} and, disentangling the lines, counted the multiplicities of the distinct diagrams that emerged. The first term can be further decomposed,%
\footnote{In our approach of directly using field components to find the shape of the orbit space, this is optional.}
\begin{equation}
\begin{split}
  \TextVCenter{\includegraphics{P33first}} 
  &= -\frac{1}{2} p_{2}^{3} + \frac{3}{2} p_{2} p_{3},
\end{split}
\end{equation}
while the second term is a new higher-order gauge invariant that is part of the polynomial basis and cannot be expressed in terms of lower-order invariants. We also have
\begin{align}
  P_{34} &= 4 \, \TextVCenter{\includegraphics{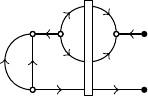}}
  = \frac{2}{3} \left( 2 \, \TextVCenter{\includegraphics{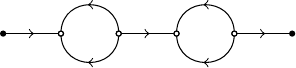}} \, 
  + 4 \, \TextVCenter{\includegraphics{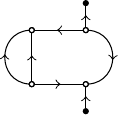}} \right)
  \notag
  \\
  &=\frac{2}{3} \left( - p_{1} p_{2}^{2} + p_{1} p_{3} + 2 p_{2} p_{4} + 4 \, \TextVCenter{\includegraphics{P34snd}} \right),
  \\
  P_{35} &= 2 \Re \, \TextVCenter{\includegraphics{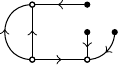}},
\\
  P_{44} &= 2 \,\TextVCenter{\includegraphics{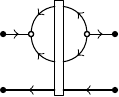}} \, + \, 2 \, \TextVCenter{\includegraphics{P44last}}
  \notag
  \\
  &= \frac{2}{3!} \left(2 \, \TextVCenter{\includegraphics{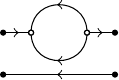}} \, + 4 \,\TextVCenter{\includegraphics{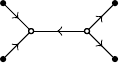}} \right) + 2 \, \TextVCenter{\includegraphics{P44last}}
  \notag
  \\
  &= \frac{2}{3} \left(p_{1} p_{4} + 2 \, \TextVCenter{\includegraphics{P44snd}}\right) 
  - p_{1} p_{2}^{2} + p_{1} p_{3} + 2 p_{2} p_{4},
\end{align}
\begin{align}
  P_{45} &= \frac{1}{2} \, \TextVCenter{\includegraphics{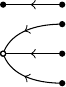}} + 
  \frac{1}{2} \, \TextVCenter{\includegraphics{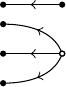}}
   +  3 \Re \TextVCenter{\includegraphics{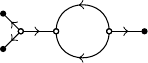}} 
  = p_{1} p_{5} + 3 \Re \TextVCenter{\includegraphics{P45last}},
\\
  P_{55} &= \frac{1}{2} \, \TextVCenter{\includegraphics{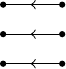}} \, + \frac{9}{2} \TextVCenter{\includegraphics{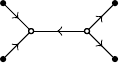}} = \frac{1}{2} p_{1}^{3} + \frac{9}{2} \TextVCenter{\includegraphics{P55last}}.
\end{align}

\subsection{Case of the $Y = 1/2$ quadruplet}
\label{sec:pmatrix:y:1:2}

In the case of the $Y = 1/2$ quadruplet, only the invariant $p_{5}$ differs from the $Y = 3/2$ case, so only the fifth column (and row) of the $P$-matrix are different. Let us denote the Levi-Civita tensor as
\begin{equation}
  \epsilon^{ij} = \TextVCenter{\includegraphics{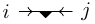}} = - \TextVCenter{\includegraphics{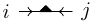}},
  \qquad 
   \epsilon_{ij} = \TextVCenter{\includegraphics{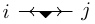}}
   = -\TextVCenter{\includegraphics{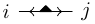}},
\end{equation}
where the triangle is needed to keep track on the orientation of legs, since $\epsilon_{ij} = -\epsilon_{ji}$ and $\epsilon^{ij} = -\epsilon^{ji}$. With this convention, we can rotate also birdtrack diagrams involving the Levi-Civita tensor with impunity.

The invariant $p_{5}$ for the $Y = 1/2$ quadruplet is given by
\begin{equation}
  p_{5} = \frac{1}{2} (\phi^{*ijk} H_{i} H_{j} \epsilon_{kl} H^{*l} + H_{l} \epsilon^{lk} H^{*i} H^{*j} \phi_{ijk})  =  \frac{1}{2} \left( \TextVCenter{\includegraphics{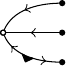}} + \TextVCenter{\includegraphics{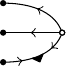}} \right),
\end{equation}
 The gradient of $p_{5}$ with respect to the field vector \eqref{eq:field:vector} is
\begin{equation}
  \frac{\partial{p_{5}}}{\partial \Phi} = \frac{1}{2}
  \left( \; \TextVCenter{\includegraphics{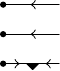}}, \; 
  \TextVCenter{\includegraphics{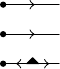}}, \;
  2 \,\TextVCenter{\includegraphics{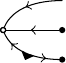}} +  \TextVCenter{\includegraphics{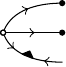}}, \;
  2 \, \TextVCenter{\includegraphics{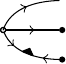}} +  \TextVCenter{\includegraphics{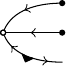}} \; \right).
\end{equation}

We have
\begin{align}
  P_{15} &=  \frac{3}{2} \left( \TextVCenter{\includegraphics{p5y12}} + \TextVCenter{\includegraphics{p5conjy12}} \right) = 3 p_{5},
  \\
  P_{25} &=  \frac{1}{2} \left( \TextVCenter{\includegraphics{p5y12}} + \TextVCenter{\includegraphics{p5conjy12}} \right) = p_{5},
\end{align}
which have the same form as in the $Y = 3/2$ case. For the rest, we obtain
\begin{equation}
  P_{35} = 2 \Re \TextVCenter{\includegraphics{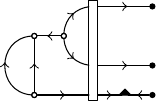}} 
  = \frac{1}{3} \Re \left( \; 2 \TextVCenter{\includegraphics{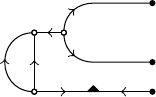}} \; 
  + 4 \TextVCenter{\includegraphics{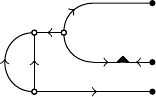}} \; \right),
\end{equation}
\begin{equation}
  P_{45} = \frac{2}{3} p_{1} p_{5} + 2 \Re \; \TextVCenter{\includegraphics{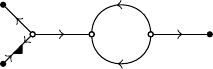}}
  +\Re \; \TextVCenter{\includegraphics{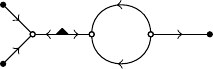}},
\end{equation}
\begin{equation}
  P_{55} = \frac{1}{6} \, \TextVCenter{\includegraphics{p1cubed}} \; +  2 \; \TextVCenter{\includegraphics{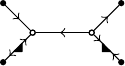}} 
  + 2 \Re \; \TextVCenter{\includegraphics{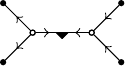}} 
  + \frac{1}{2} \TextVCenter{\includegraphics{P55last}},
\end{equation}
where in the last term we used that $\epsilon_{ij} \epsilon^{jk} = \delta^{k}_{i}$ or $\includegraphics{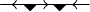} = \includegraphics{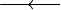}$ and $H^{* i} \epsilon_{ij} H^{* j} = \TextVCenter{\includegraphics{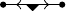}} = 0$.

\acknowledgments

We would like to thank Gauthier Durieux, Matthew McCullough, Ennio Salvioni, Hardi Veerm\"{a}e and Luis Lavoura for discussions, and Niko Koivunen and Luca Marzola for reading the manuscript and useful comments. This work was supported by the Estonian Research Council grants PRG803, RVTT3 and RVTT7, and the CoE program TK202 ``Fundamental Universe''.

\bibliographystyle{JHEP}
\bibliography{SU2_quadruplet_orbit_space}

\end{document}